\documentclass[%
reprint,
superscriptaddress,
nofootinbib,
amsmath,amssymb,
aps,
showkeys,
]{revtex4-1}
\usepackage{graphicx}
\usepackage{dcolumn}
\usepackage{bm}
\usepackage[%
bookmarksnumbered,
bookmarksopen,
colorlinks,
citecolor=blue,
linkcolor=blue,
]{hyperref} 

\def\esym{$E_{sym}(\rho)$~}
\def\rpi {$\pi^-/\pi^+$~}
\def\es0{$E_{sym}(\rho_0)$~}
\begin{document}


\title{Examination of an isospin-dependent single-nucleon momentum distribution for the isospin-asymmetric nuclear matter in heavy-ion collisions}

\author{Gao-Feng Wei}\email[Corresponding author. E-mail: ]{wei.gaofeng@gznu.edu.cn}
\affiliation{School of Physics and Electronic Science, Guizhou Normal University, Guiyang 550025, China}
\affiliation{Guizhou Provincial Key Laboratory of Radio Astronomy and Data Processing, Guizhou Normal University, Guiyang 550025, China}
\author{Qi-Jun Zhi}
\affiliation{School of Physics and Electronic Science, Guizhou Normal University, Guiyang 550025, China}
\affiliation{Guizhou Provincial Key Laboratory of Radio Astronomy and Data Processing, Guizhou Normal University, Guiyang 550025, China}
\author{Xin-Wei Cao}
\affiliation{School of Mechanical and Material Engineering, Xi'an University of Arts and Sciences, Xi'an 710065, China}
\author{Zheng-Wen Long}
\affiliation{College of Physics, Guizhou University, Guiyang 550025, China}


\begin{abstract}
	
Within a transport model using as the input nucleon momentum profiles from a parameterized isospin-dependent single-nucleon momentum distribution with a high momentum tail induced by short-range correlations, we employ the $^{197}$Au + $^{197}$Au collisions at 400 MeV/nucleon to examine on one hand effects of the short-range correlations on the pion and flow observables in probing the nuclear symmetry energy, and on the other hand how reliable are this isospin-dependent single-nucleon momentum distribution as well as the corresponding parameter settings. Besides significant effects of the short-range correlations on the pion and flow observables are observed, we also find that the theoretical simulations of $^{197}$Au + $^{197}$Au collisions with this momentum distribution using two sets of parameters extracted from the experimental analysis and the self-consistent Green's function prediction, respectively, can reproduce the neutron elliptic flows of the FOPI-LAND experiment and the \rpi ratios of the FOPI experiment under the symmetry energy setting in a certain range. Therefore, we conclude that this parameterized isospin-dependent single-nucleon momentum distribution is reliable for the isospin-asymmetric nuclear matter, correspondingly, two sets of parameters extracted from both the experimental analysis and the self-consistent Green's function prediction can not be ruled out according to the available experimental information at present.

\end{abstract}

\keywords{single-nucleon momentum distribution; isospin-asymmetric nuclear matter; short-range correlations; heavy-ion collisions, symmetry energy}
\maketitle


\section{introduction}\label{introduction}

The determination of the equation of state (EoS) of dense isospin-asymmetric nuclear matter (IANM) has always been a fascinating problem in nuclear physics and nuclear astrophysics due to its vital importance
in studying the structure of radioactive nuclei and the evolution of compact stars, and thus attracted much attention in the past few decades, see, e.g., Refs.~\cite{Lat12,Hor14} for recent comprehensive reviews. Nevertheless, the predictions on the EoS of dense IANM, especially its density-dependent symmetry energy term are still discrepant even controversial at the suprasaturation density~\cite{Xiao09,Xie13,Feng10}, although many isospin signals have been proposed aiming to detect the EoS of dense IANM, see, e.g., Refs.~\cite{Lat12,Hor14,Zhang18,Yan19}. This is mainly because the isovector part of nuclear interactions is much weaker than its isoscalar part, and thus these isospin signals can usually be interfered with by other factors in theoretical simulations and experimental measurements. Therefore, some attempts on strategic studies~\cite{Tsang17} and covariance analyses~\cite{Zhang15} of these isospin observables as well as some comparative projects~\cite{SEP} between different theoretical communities have been carried out to understand the origin of these discrepancies. Actually, besides these attempts, some studies on the deficiency of mechanism itself in theoretical simulations are also the correct direction to the solution of these discrepancies, for instance, the pion potential~\cite{Hong14,Guo15a,Feng18} and the $\Delta$ isovector potential~\cite{Bao15a,Guo15b} have been confirmed that they can all interfere with the sensitivity of the \rpi ratio in probing the nuclear symmetry energy using heavy-ion collisions (HICs).

On the other hand, the momentum distribution of nucleons in a nuclear system, as the direct reflection of strong interactions at short distances, has always been a long-standing interest in nuclear physics~\cite{Subedi08,Ciofi15}. In particular, the discovery of correlated nucleons pairs in a $^{12}$C nucleus in high energy electron scattering experiments at the Jefferson Laboratory (JLab)~\cite{Subedi08} arouses the higher enthusiasm in studying the momentum distribution of nucleons and their short-range correlations (SRCs) in the past decades, see, e.g., Refs.~\cite{Ciofi15,LiBA18} for comprehensive reviews. Qualitatively, people have already gained some general knowledge on the nucleon momentum distribution (NMD), i.e., the tensor component of nuclear interactions can usually push a few nucleons from low momentum to high momentum, leading to a high momentum tail (HMT) above the nucleon Fermi momentum $k_{F}$ and a corresponding low momentum depletion (LMD) below the $k_{F}$ in the NMD~\cite{Ciofi15,LiBA18}. Moreover, a qualitative consensus that the $np$ dominance of SRC pairs in the HMT has been confirmed by various theoretical and experimental investigations~\cite{Ohen14,Wir14,Sar14,Ryck15}. Quantitatively, the experimental analyses at the JLab suggest that approximately 20\% of nucleons are in the HMT in a nucleus from light $^{12}$C~\cite{Subedi08} even to heavy $^{208}$Pb~\cite{Ohen14,Ohen18}; Also, the systematic analyses of these results in experiments at the JLab indicate that the fraction of nucleons in the HMT is approximately 25\% in the symmetric nuclear matter (SNM) at the saturation density $\rho_{0}$~\cite{Ohen15}, it is however, the theoretical calculations begin to deviate significantly from this suggested fraction for the HMT in the SNM at $\rho_{0}$. For example, the self-consistent Green's function (SCGF) approach employing the Av18 interaction predicts only 11-13\% of nucleons in the HMT for the SNM at $\rho_{0}$ but a higher fraction 4-5\% of nucleons in the HMT for the pure neutron matter (PNM) at $\rho_{0}$~\cite{Rios09}, while the Bruckner-Hartree-Fock calculations go so far as to suggest a wide ranges for nucleons in the HMT in the SNM at $\rho_{0}$ from about 10\% using the N3LO450 interaction to about 20\% using the Av18, Paris, or Nij93 interactions~\cite{LiZH16}. 

Recently, guided by the earlier studies aforementioned, a parameterized isospin-dependent single NMD with a HMT induced by SRCs, i.e.,
\begin{eqnarray}\label{nk}
n^{J}_{\bf k}(\rho,\delta)=\left\{%
\begin{array}{ll}
{ \Delta}_{J}+\beta_{J}I{\big(}|{\bf k}|/k^{J}_{F}{\big)}, & \hbox{$0<|{\bf k}| \leq k^{J}_{F}$,} \\
C_{J}{\big(}k^{J}_{F}/|{\bf k}|{\big)}^{4}, & \hbox{$k^{J}_{F} < |{\bf k}| \leq \phi_{J}k^{J}_{F}$}, \\
\end{array}%
\right.
\end{eqnarray}
is proposed for the IANM in Refs.~\cite{Cai15,LiBA18}.  Here, the parameters ${\Delta}_{J}$ and $\beta_{J}$ characterize, respectively, the depletion of the Fermi sphere at zero momentum, and the strength of momentum-dependent depletion through the momentum-dependent function $I(|{\bf k}|/k^{J}_{F})\approx(|{\bf k}|/k^{J}_{F})^{2}$ in the range of $0<|{\bf k}| \leq k^{J}_{F}$ with $k^{J}_{F}=(3\pi^2\rho_{J})^{1/3}$ and the index $J$ denotes protons or neutrons, while the parameters $C_{J}$ and $\phi_{J}$ are the amplitude and cutoff value of high momentum distribution, respectively; All of these four parameters can be expressed in isospin asymmetry $\delta$ in the general form of $Y_{J}=Y_{0}(1+Y_{1}\tau^{J}_{3}\delta)$ with the $\tau^{n}_{3}$ is 1 for neutrons and the $\tau^{p}_{3}$ is -1 for protons. Nevertheless, it should be mentioned that the normalization condition $[2/(2\pi)^{3}]\int_{0}^{\infty}n^{J}_{\bf k}(\rho,\delta){\rm d}{\bf k}$=$\rho_{J}$=$(k^{J}_{F})^{3}/3\pi^{2}$ requires that only three of these four parameters are independent, it is therefore one can use the parameters $C_{J}$, $\phi_{J}$ and $\beta_{J}$ as the independent ones as that in Refs.~~\cite{Cai15,LiBA18}. Moreover, to cover approximately the recent experimental findings~\cite{Ohen14,Ohen15} and the SCGF calculations~\cite{Rios09}, the authors of Refs.~\cite{Cai15,LiBA18} extracted two sets of parameters. One is the HMT-SCGF parameter, i.e., $C_{0}=0.121$, $C_{1}=-0.01$, $\phi_{0}=1.49$, $\phi_{1}=-0.25$ and $\beta_{0}=\beta_{1}=0$, which leads to a fraction of total nucleons approximately 12.6\% in the HMT in finite $^{197}$Au nucleus at $\rho_{0}$, correspondingly, the fractions of high-momentum protons and neutrons over total protons and neutrons are approximately 15\% and 11\%, respectively; Another is the HMT-exp. parameter, i.e., $C_{0}=0.161$, $C_{1}=-0.25$, $\phi_{0}=2.38$, $\phi_{1}=-0.56$, $\beta_{0}=-0.27$ and $|\beta_{1}|\leq 1$, which leads to a fraction of total nucleons approximately 25.2\% in the HMT in finite $^{197}$Au nucleus at $\rho_{0}$, and the individual fractions of high-momentum protons and neutrons are approximately 30\% and 22\%, respectively. For more details about this isospin-dependent NMD, we refer readers to see Refs.~\cite{Cai15,LiBA18}. 

\begin{figure}[t]
	\includegraphics[width=0.9\columnwidth]{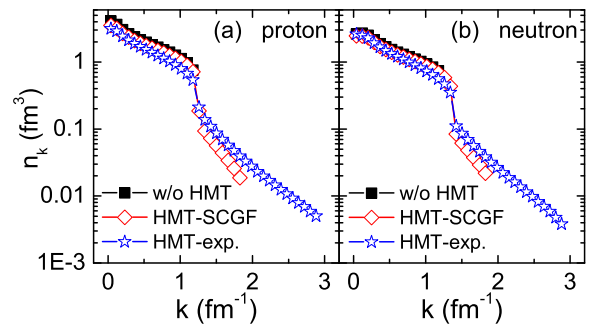}
	\caption{(Color online) Momentum distributions of protons (a) and neutrons (b) in the initial $^{197}$Au nucleus with and without the consideration of SRCs. The normalization condition $[2/(2\pi)^{3}]\int_{0}^{\infty}n^{J}_{\bf k}(\rho,\delta){\rm d}{\bf k}$=$\rho_{J}$=$(k^{J}_{F})^{3}/3\pi^{2}$ is used.} \label{mdis}
\end{figure}

\section{The Model}\label{Model}
\subsection{Initialization of the model}\label{INIT}
The present study is carried out in the framework of an isospin- and momentum-dependent Boltzmann-Uehling-Uhlenbeck (IBUU) transport model \cite{IBUU}. As the first step, the startup of the IBUU model is the initialization of colliding nuclei in the coordinate and momentum spaces. 
Certainly, to initialize the finite $^{197}$Au nucleus in the momentum space, we have transformed above NMD for the  IANM into that for the finite $^{197}$Au nucleus using the local-density approximation~\cite{Vive04}. The specific distributions generated from the initialization in our reaction model are shown in Fig.~\ref{mdis}.  Consistent with the previous theoretical results~\cite{Sar14,Rios14,Wir14,Ryck15} and the recent experimental evidence~\cite{Ohen18}, a higher fraction of high-momentum protons than that of neutrons in a neutron-rich nuclear system is resulting from the $np$ dominance of SRC pairs in the HMT. Moreover, except for the observed LMD below the $k_{F}$ and the corresponding HMT above the $k_{F}$, the nucleon momentum profiles generated from this isospin-dependent NMD are very similar to those without the consideration of SRCs, and thus preliminarily indicating the feasibility of initialization using this isospin-dependent NMD in the momentum space. Certainly, the reliability of specific parameter settings based on this distribution still needs to be checked as shown in the following parts. Also, because the fraction of high-momentum nucleons calculated with the HMT-exp. parameter is higher than that in calculations with the HMT-SCGF parameter, more apparent LMD below the $k_{F}$ and the corresponding HMT above the $k_{F}$ can be seen in calculations with the HMT-exp. parameter. Therefore, some sensitive observables of the NMD are expected to reflect these differences in $^{197}$Au+$^{197}$Au collisions. Moreover, we have also checked effects of the value $\beta_{1}$ in the allowed range on the NMD in $^{197}$Au nucleus and found that the NMD of the $^{197}$Au nucleus is less influenced by the value of $\beta_{1}$ parameter, it is thus we randomly take the value for the $\beta_{1}$ in the allowed range in this study.

\subsection{Interaction used in the model}\label{interaction}
As far as the nuclear interaction used in the present study, we take the form similar to our recent work~\cite{Wei18a,Wei18b} because this nuclear interaction has considered more reasonable details including distinguishing the density dependencies of in-medium $nn$, $pp$ and $np$ interactions in the effective many-body force term~\cite{Chen14b} as well as fitting the high-momentum behaviors of the nucleon optical potential extracted from nucleon-nucleus-scattering experiments~\cite{LXH15}. Specifically, the nuclear interaction used in this study is expressed as
\begin{eqnarray}
U(\rho,\delta ,\vec{p},\tau ) &=&A_{u}(x)\frac{\rho _{-\tau }}{\rho _{0}}%
+A_{l}(x)\frac{\rho _{\tau }}{\rho _{0}}+\frac{B}{2}{\big(}\frac{2\rho_{\tau} }{\rho _{0}}{\big)}^{\sigma }(1-x)  \notag \\
&+&\frac{2B}{%
	\sigma +1}{\big(}\frac{\rho}{\rho _{0}}{\big)}^{\sigma }(1+x)\frac{\rho_{-\tau}}{\rho}{\big[}1+(\sigma-1)\frac{\rho_{\tau}}{\rho}{\big]}
\notag \\
&+&\frac{2C_{l }}{\rho _{0}}\int d^{3}p^{\prime }\frac{f_{\tau }(%
	\vec{p}^{\prime })}{1+(\vec{p}-\vec{p}^{\prime })^{2}/\Lambda ^{2}}
\notag \\
&+&\frac{2C_{u }}{\rho _{0}}\int d^{3}p^{\prime }\frac{f_{-\tau }(%
	\vec{p}^{\prime })}{1+(\vec{p}-\vec{p}^{\prime })^{2}/\Lambda ^{2}},
\label{MDIU}
\end{eqnarray}%
and the corresponding parameters $A_{l}(x)$ and $A_{u}(x)$ are expressed in forms of
\begin{eqnarray}
A_{l}(x)&=&A_{l0} - \frac{2B}{\sigma+1}\big{[}\frac{(1-x)}{4}\sigma(\sigma+1)-\frac{1+x}{2}\big{]},  \\
A_{u}(x)&=&A_{u0} + \frac{2B}{\sigma+1}\big{[}\frac{(1-x)}{4}\sigma(\sigma+1)-\frac{1+x}{2}\big{]}.
\end{eqnarray}
The parameter $x$ embedded in above expressions is to mimic the slope value $L\equiv{3\rho({dE_{sym}}/d\rho})$ of nuclear symmetry energy at $\rho_{0}$ predicted by various many-body theories without changing the value of nuclear symmetry energy $E_{sym}(\rho)$ at $\rho_{0}$ and any properties of the SNM. Generally, without the consideration of correlations in a nuclear system, the kinetic part of nuclear symmetry energy is calculated from the free Fermi gas model as $E^{kin}_{sym}(\rho)=8\pi p^{5}_{f}/{9m h^{3}\rho}{\approx}12.5(\rho/\rho_{0})^{2/3}$ with $p_{f}=\hbar (3\pi^{2}\rho/2)^{1/3}$. Nevertheless, considering that the SRCs reduce significantly the $E^{kin}_{sym}(\rho)$ according to some microscopic calculations~\cite{Vida11,Rios14} and experimental studies~\cite{Ohen14,Ohen15}, we therefore should modify this expression. Actually, as was indicated in Ref.~\cite{Rios14}, what we can identify as correlation driven is the reduction of $E^{kin}_{sym}(\rho)$,  it is therefore the $E^{kin}_{sym}(\rho)$ is the sole criterion that can guide one to incorporate the SRC effects into nuclear effective interactions  phenomenologically. To this end, we readjust the parameters of nuclear interactions to phenomenologically incorporate the SRC effects into nuclear effective interactions for the SRC scenario. Moreover, considering that only minority of nucleons are SRC correlated, we therefore assume that the kinetic symmetry energy also holds for the 2/3 regularity with respect to densities. Also, according to a microscopic calculation~\cite{Vida11}, the symmetry energy at $\rho_{0}$ is almost from its potential part. It is for this reason we employ the expression $E^{kin}_{sym}(\rho)=12.5{\big[}(\rho/\rho_{0})^{2/3}-1{\big]}$ for the kinetic symmetry energy to meet this demand for the SRC scenario similar to the previous study~\cite{Yong16}. Using empirical constraints on the properties of symmetric and/or asymmetric nuclear matter at $\rho_{0}=0.16$ fm$^{-3}$, see, e.g., Ref.~\cite{Wei20} for detailed constraints used in this study, 
we have fitted parameters of nuclear interactions, i.e., $A_{l0}$, $A_{u0}$, $B$, $C_{l}$, $C_{u}$, $\sigma$, and $\Lambda$, for the scenarios without and with SRCs, respectively. Specifically, The values of two sets parameters denoted as "w/o HMT" and "with HMT" are shown in Table~\ref{tab1}. 

\begin{figure}[t]
	\includegraphics[width=0.9\columnwidth]{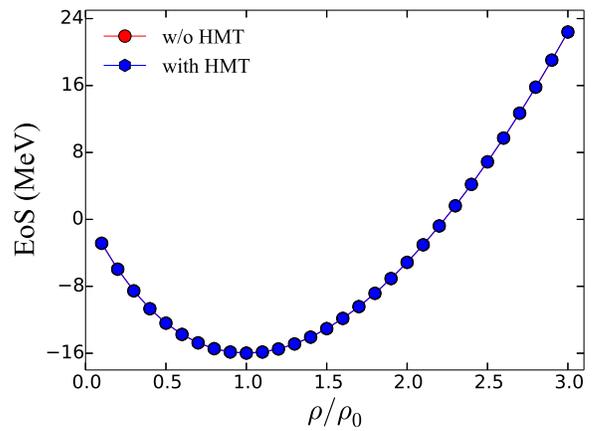}
	\caption{(Color online) EoS of symmetric nuclear matter calculated with (blue solid hexagon) and without (Red solid circle) the consideration of SRCs.} \label{EoS}
\end{figure}
\begin{center}
	\begin{table}[htb]
		\caption{{\protect\small The parameters used in the present study.}} \label{tab1}
		\begin{tabular}{c   c   c}
			\hline
			\hline
			\quad parameters & \quad w/o HMT  & \quad with HMT \\
			\hline
			$A_{l0}$~(MeV) \quad & $-96.963$ \quad & $-96.963$ \\
			$A_{u0}$~(MeV) \quad & $-36.963$ \quad & $-36.963$ \\
			$C_{l}$~(MeV) \quad & $-40.820$ \quad & $-24.719$ \\
			$C_{u}$~(MeV) \quad & $-119.368$ \quad & $-135.469$ \\
			$B$~(MeV) \quad & $141.963$ \quad & $141.963$ \\
			$\sigma$ \quad & $1.2652$ \quad & $1.2652$ \\
			$\Lambda/p_{f}$ \quad & $2.424$ \quad & $2.424$ \\
			\hline
		\end{tabular}%
	\end{table}
\end{center}

Obviously, the isoscalar potentials $U_{0}({\rho},p)$ under the consideration of SRCs are completely identical with those without the consideration of SRCs at either low densities or high densities due to only the symmetry energy is used as the criterion of the correlation-driven effects. As a result, the corresponding EoS of SNM is completely identical to each other in these scenarios as shown in Fig.~\ref{EoS}. Actually, as indicated in Eq.~({\color{blue}{34}}) of Ref.~\cite{Chen14b}, the EoS of SNM is involved in the combination of parameters $C_{l}$ and $C_{u}$ instead of their individuals, i.e., $C_{l}+C_{u}$, the SRCs naturally will not affect the EoS of SNM at either low densities or high densities. Nevertheless, with the consideration of SRCs, the symmetric potentials will begin to deviate significantly from those without the consideration of SRCs. Naturally, through decomposing the single-nucleon potential in Eq. (\ref{MDIU}), i.e., $U_{p,n}(\rho,\delta,p){\approx}U_{0}(\rho,p)+U_{sym}(\rho,p)(\tau\delta)$ with $\tau=1$ for neutrons and $-1$ for protons, one can find that the nuclear interaction in Eq. (\ref{MDIU}) employing the "with HMT" parameter is essentially different from that with the "w/o HMT" parameter although an identical expression of nuclear interactions in form is used in these two different scenarios. 
Certainly, in principle, for the SRC scenarios but different initialization, i.e., the HMT-SCGF parameter and HMT-exp. parameter, we should exactly fit different nuclear interaction parameters according to the different fractions of SRC nucleons instead of an identical interaction parameter, i.e., "with HMT" parameter. However, the quantitative results regarding this issue are still inconclusive according to present microscopic calculations~\cite{Rios14,Vida11}; In other words,  due to what we can identify as correlation driven is the reduction of $E^{kin}_{sym}(\rho)$~\cite{Rios14},  it is therefore the $E^{kin}_{sym}(\rho)$ is the sole criterion that can guide one to incorporate the SRC effects into nuclear effective interactions  phenomenologically.
On the other hand, considering the fact that the shapes of NMD for these two cases are similar to each other, their differences are only the initial maximum momentum and the specific fraction of high-momentum nucleons, it is therefore the utilization of identical nuclear interactions in form for these two cases is suitable at the level of mean field. Nevertheless, it should be emphasized that the initialization effects and thus the resulting differences of initial maximum momentum of SRC nucleons as well as their fractions in these two cases will also be incorporated into dynamics process of reactions through the momentum dependent $C_{l}$ and $C_{u}$ integral terms. 

Instead of the traditional Coulomb field used in theoretical simulations of nucleus-nucleus collisions, the Li\'{e}nard-Wiechert formulas~\cite{Ou11,Deng17} are also used to the relativistic calculations of electromagnetic interactions created by fast-moving charged particles during HICs in the present study, i.e.,
\begin{equation}\label{E-field}
e\vec{E}(\vec{r},t)=\frac{e^2}{4\pi \varepsilon_{0}}
\sum_{n}Z_{n}\frac{c^2-v^{2}_{n}}{(cR_{n}-\vec{R}_{n}\cdot \vec{v}_{n})^3}(c\vec{R}_{n}-R_{n}\vec{v}_{n}),
\end{equation}
\begin{equation}\label{M-Field}
e\vec{B}(\vec{r},t)=\frac{e^2}{4\pi \varepsilon_{0}}
\sum_{n}Z_{n}\frac{c^2-v^{2}_{n}}{(cR_{n}-\vec{R}_{n}\cdot \vec{v}_{n})^3}\vec{v}_{n}\times \vec{R}_{n},
\end{equation}
because they can also appreciably affect some isospin observables such as the \rpi ratio and the neutron-proton differential transverse flow in HICs at intermediate energies especially around 400 MeV/nucleon, see., e.g., Refs.~\cite{Wei18a,Wei18b} for more details. Moreover, to improve the accuracies of theoretical simulations of HICs, we have also considered the pion potential and the $\Delta$ isovector potential in the present study. Specifically, when the pionic momentum is higher than 140 MeV/$c$, we adopt the pion potential based on $\Delta-$hole model of the form used in Ref.~\cite{Buss04}, and when the pionic momentum is lower than 80 MeV/$c$, we use the pion potential of the form used in Refs.~\cite{Eric66,Oset88,Oset93}, while for the pionic momentum falling into the range from 80 to 140 MeV/$c$, an interpolative pion potential constructed by O. Buss in Ref.~\cite{Buss04} is used. As far as the effects of this pion potential on the isospin observables such as the \rpi ratio in HICs, we refer readers to see Ref.~\cite{Guo15a} for more details. For the $\Delta$ potential, guided by the earlier studies~\cite{Bao15a,Guo15b} and according to the decay mechanism of $\Delta$ resonances, we use an isospin-dependent $\Delta$ potential in the present study, i.e.,
\begin{eqnarray}\label{delta}
U(\Delta^{++})&=&f_{\Delta}U(p),\\
U(\Delta^{+})&=&f_{\Delta}{\big[}\frac{1}{3}U(n)+\frac{2}{3}U(p){\big]},\\
U(\Delta^{0})&=&f_{\Delta}{\big[}\frac{2}{3}U(n)+\frac{1}{3}U(p){\big]},\\
U(\Delta^{-})&=&f_{\Delta}U(n).
\end{eqnarray}
Certainly, the quantitative results of the $\Delta$ potential is still inconclusive at present, it is however, considering the fact that the depth of nucleon potential is approximately $-50$ MeV, while that of the $\Delta$ potential is empirically constrained around $-30$ MeV，
it is for this reason that we additionally introduce an identical factor $f_{\Delta}=2/3$ for them to meet this demand in spite of this factor maybe differnt for the $\Delta$ resonances with different charged states.

\section{Results and Discussions}\label{Results and Discussions}

\begin{figure}[t]
	\centerline{\includegraphics[width=\columnwidth]{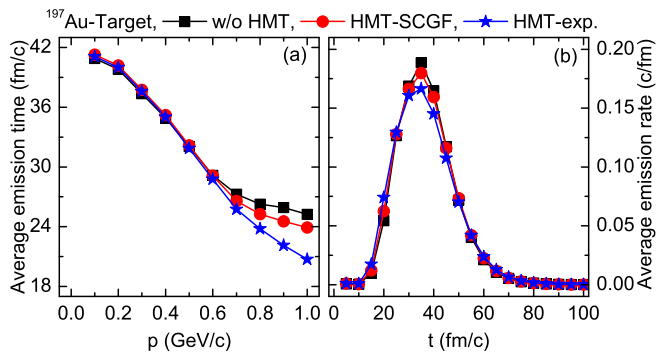}}
	\caption{(Color online) Average emission time (a) and emission rate (b) for nucleons in scenarios with and without the consideration of SRCs. The soft symmetry energy with parameter $x=1$ is used.}\label{Stable}
\end{figure}

Before presenting results of our studies, we first examine the emission time and emission rate of nucleons during collisions to check the stability of our model. To this end, we use the criterion of the last strong interaction to calculate the average emission time and emission rate for nucleons in target of $^{197}$Au + $^{197}$Au collisions without distinguishing the proton and neutron. Statically understanding, the presence of nucleons with momentum more than $k_{F}$ in the initial target will naturally cause an increase of nucleon emissions in the initial reaction stage due to the larger pressure generated in colliding region by these high-momentum nucleons as well as their violent collisions. Certainly, the effects of correlations between these nucleons will cause a corresponding reduction of nucleon emissions to compete with the former. Nevertheless, at the initial reaction stage, the former will dominate than the latter. As expected, for nucleons with momentum greater than about 600 MeV/$c$, their average emission time indeed shows rather sensitivities to the fraction of SRC nucleons as shown in panel (a) of Fig.~\ref{Stable}, i.e., compared to cases excluding and/or including fewer SRC nucleons in the target, the case including more SRC nucleons in the colliding nuclei gets the preequilibrium nucleons faster emission. Also, as shown in panel (b) of Fig.~\ref{Stable}, the nucleon emission rate before the moment about 27 fm/$c$ shows more larger value in SRC scenarios, in particular for that with the HMT-exp. parameter, this implies that more SRC nucleons in the target cause more nucleons emitting in this period. Actually, at the early reaction stage, as the projectile gradually approaches and then compresses the target, nucleons in target with momentum more than $k_{F}$ have larger probabilities to be accelerated into the region of momentum greater than about 600 MeV/c, this is the reason we see in panel (b) of Fig.~\ref{Stable} the larger emission rate before the moment about 27 fm/$c$ in collisions for the scenario with more SRC nucleons in the colliding nuclei. Certainly, for the SRC scenario, with the emission of more preequilibrium nucleons at the early reaction stage, the remainders naturally are less than those in collisions without the consideration of SRCs. Consequently, for the later reaction stages, the nucleon emission rate in collisions without SRCs will dominate than that in collisions with SRCs. Indeed, for the nucleon emission after about 27 fm/$c$, the smaller emission rate is seen in the scenario with more SRC nucleons in the colliding nuclei as shown in panel (b) of Fig.~\ref{Stable}. 
Moreover, before and/or even at 15 fm/$c$, we can find that the nucleon emission rate is not more than 2.5\%, the stability of ground states at this level therefore guarantees the statistical results obtained here good enough.
\begin{figure}[t]
	\centerline{\includegraphics[width=0.9\columnwidth]{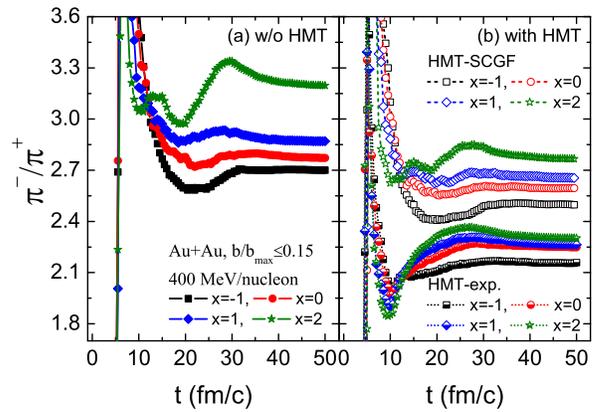}}
	\caption{(Color online) Time evolutions of the $(\pi^{-}/\pi^{+})_{\rm like}$ ratios in central $^{197}$Au + $^{197}$Au collisions at 400 MeV/nucleon with the symmetry energy ranging from the superhard with $x=-1$ to the supersoft with $x=2$. The panels (a) and (b) are the results in scenarios without and with the consideration of SRCs, respectively. } \label{rpi}
\end{figure}

Now, let's check the influences of SRCs on the final reaction products in $^{197}$Au + $^{197}$Au collisions. First, we examine effects of the SRCs on the pion observable. To this end, the dynamic ratio (\rpi)$_{\rm like}$ defined as
\begin{equation}
(\pi^{-}/\pi^{+})_{\rm like}\equiv\frac{\pi^{-}+\Delta^{-}+\frac{1}{3}\Delta^{0}}{\pi^{+}+\Delta^{++}+\frac{1}{3}\Delta^{+}},
\end{equation}
can be used to check the effects of SRCs on the dynamic production of pions. Certainly, the dynamic ratio (\rpi)$_{\rm like}$ will naturally become the free \rpi ratio at the end of reactions due to all the $\Delta$ resonances will decay into nucleons and pions. Shown in Fig.~\ref{rpi} are the time evolutions of dynamic ratios (\rpi)$_{\rm like}$ in central $^{197}$Au + $^{197}$Au collisions at 400 MeV/nucleon with and without the consideration of SRCs. First, it can be seen that, without the consideration of SRCs, the dynamic ratio (\rpi)$_{\rm like}$ during reactions and thus the \rpi ratio at the end of reactions are larger on the whole than those with the consideration of SRCs, regardless of the HMT-SCGF parameter or the HMT-exp. parameter is used in calculations. Second, consistent with the established systematics for pion productions~\cite{FOPIb}, it can be found that the \rpi ratio is indeed sensitive to the density-dependent nuclear symmetry energy \esym. 
Nevertheless, we can also observe a decreased sensitivity of the \rpi ratio to the nuclear symmetry energy with the consideration of SRCs, and this phenomenon is more apparent for the case with the HMT-exp. parameter. Actually, this observation can be easily understood within the reached consensus about the SRC nucleons in the HMT. That is, due to the $np$ dominance of SRC-nucleon pairs in the HMT~\cite{Ohen14,Wir14,Sar14,Ryck15}, the higher fraction of SRC nucleons in the HMT naturally leads to more $np$ pairs in the reaction system, and thus causes a more reduction of isospin asymmetry of the reaction system. Naturally, the dynamic ratio (\rpi)$_{\rm like}$ during reactions and thus the final \rpi ratio show a decreased sensitivity to the nuclear symmetry energy. 
Nevertheless, to understand the first observation, we need to look at the dynamic multiplicities of $\pi^{-}$ and $\pi^{+}$ in a quantitative manner in the following.
\begin{figure}[t]
	\centerline{\includegraphics[width=0.9\columnwidth]{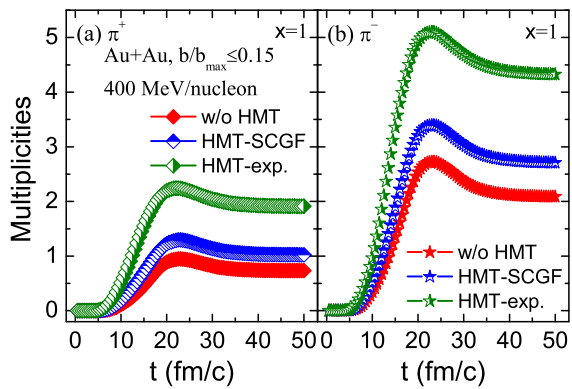}}
	\caption{(Color online) Multiplicities evolutions of dynamic $\pi^{+}$ (a) and $\pi^{-}$ (b) in central $^{197}$Au + $^{197}$Au collisions at 400 MeV/nucleon in scenarios with and without the consideration of SRCs. The soft symmetry energy with parameter $x=1$ is used.} \label{mul}
\end{figure}

\begin{figure}[t]
	\centerline{\includegraphics[width=0.9\columnwidth]{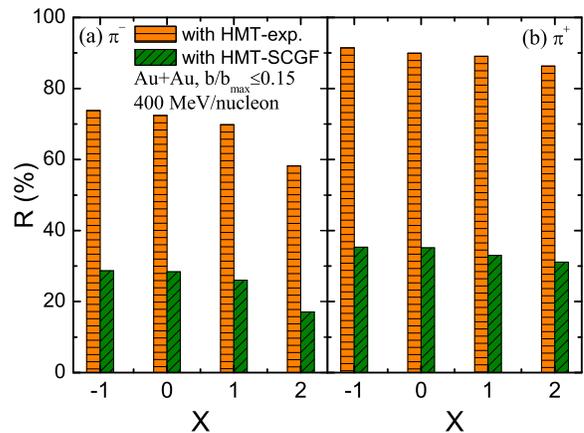}}
	\caption{(Color online) Relative effects of the SRCs on the final multiplicities of $\pi^{-}$ (a) and $\pi^{+}$ (b) in central $^{197}$Au + $^{197}$Au collisions at 400 MeV/nucleon.} \label{refcs}
\end{figure}

Shown in Fig.~\ref{mul} are the multiplicities evolutions of dynamic $\pi^{-}$ and $\pi^{+}$ with a certain symmetry energy with parameter $x=1$ in the same reaction, which are determined by $\pi^{-}+\Delta^{-}+\frac{1}{3}\Delta^{0}$ and $\pi^{+}+\Delta^{++}+\frac{1}{3}\Delta^{+}$ according to the decay mechanism of $\Delta$ resonances. It is obvious to see that more $\pi^{-}$ and $\pi^{+}$ are produced in scenarios with than without the consideration of SRCs. Consequently, the dynamic ratio (\rpi)$_{\rm like}$ during reactions and thus the final \rpi ratio may alter to different extent according to the specific fraction of SRC nucleons. More specifically, according to the following formula
\begin{equation}
R\equiv\frac{|(\pi^{\pm})_{\rm with}-(\pi^{\pm})_{\rm w/o}|}{{\big[}(\pi^{\pm})_{\rm with}+(\pi^{\pm})_{\rm w/o}{\big]/2}}\times100\%,
\end{equation}
we can calculate the relative effects of SRCs on the final multiplicities of $\pi^{-}$ and $\pi^{+}$, the corresponding results employing the HMT-exp. parameter and the HMT-SCGF parameter are shown in Fig.~\ref{refcs}, respectively. It is seen that, regardless of the HMT-SCGF parameter or the HMT-exp. parameter is used in calculations, the increment of $\pi^{+}$ is more than that of $\pi^{-}$, leading to a corresponding reduction of the \rpi ratio. This is why we see in Fig.~\ref{rpi} the \rpi ratio becomes small on the whole in scenarios with the consideration of SRCs. One may suspect why the increment of $\pi^{+}$ is more than that of $\pi^{-}$ in reactions for a certain parameter settings. Actually, just as we discussed above, because of the $np$ dominance of SRC-nucleon pairs and thus more larger probabilities of protons than neutrons in the HMT in neutron-rich systems, more energetic proton-proton collisions can cause more $\pi^{+}$ productions. Naturally, compared to the case with the HMT-SCGF parameter, the case with the HMT-exp. parameter can lead to more high-momentum nucleons in the colliding nuclei and thus more increment of $\pi^{+}$ than $\pi^{-}$ as well as more $np$ pairs, it is for this reason the corresponding \rpi ratio has more smaller values and weaker sensitivity to the nuclear symmetry energy.

\begin{figure}[t]
	\centerline{\includegraphics[width=0.9\columnwidth]{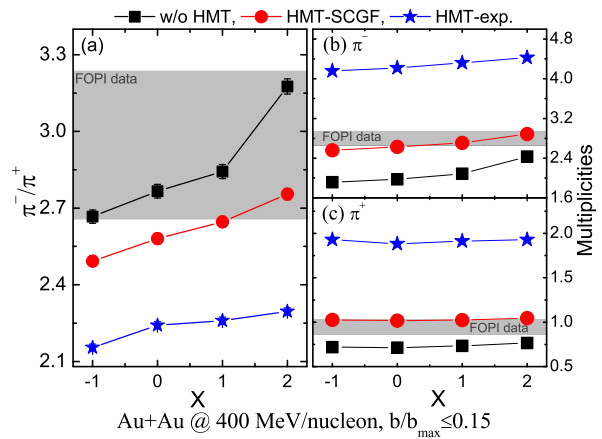}}
	\caption{(Color online) Final multiplicities of $\pi^{-}$ and $\pi^{+}$ and the corresponding \rpi ratios compared with those from the FOPI experiment in central $^{197}$Au + $^{197}$Au collisions at 400 MeV/nucleon without and with the consideration of SRCs.} \label{com1}
\end{figure}

Before checking the effects of SRCs on the collective flow of nucleons, let's first compare the final multiplicities of $\pi^{-}$ and $\pi^{+}$ as well as their \rpi ratios with the available data to examine the reliability of this isospin-dependent single NMD as well as the corresponding two sets of parameters for the IANM. Shown in panels (b) and (c) of Fig.~\ref{com1} are the final multiplicities of $\pi^{-}$ and $\pi^{+}$ generated in central $^{197}$Au + $^{197}$Au collisions for the scenarios with SRCs as well as those from the FOPI experiment~\cite{FOPIb}. As a comparison, we also show the corresponding results of central $^{197}$Au + $^{197}$Au collisions for the scenario without SRCs. Obviously, without the consideration of SRCs, the multiplicities of both $\pi^{-}$ and $\pi^{+}$ are much underestimated on the whole in calculations with all $x$ parameters used here; On the contrary, the multiplicities of both $\pi^{-}$ and $\pi^{+}$ are much overestimated in calculations using the HMT-exp. parameter; While for the case using the HMT-SCGF parameter, the theoretical simulations of $^{197}$Au + $^{197}$Au collisions can reproduce the multiplicities of both $\pi^{-}$ and $\pi^{+}$ of the FOPI experiment with the symmetry energy parameter $x$ setting approximately in the range from 0 to 2. Certainly, to reduce the systematic errors in probing the symmetry energy using the multiplicities of charged pions, their ratios \rpi are usually used as more effective observable of the symmetry energy in HICs at intermediate energies. To this end, we show in panel (a) of Fig.~\ref{com1} the ratios \rpi of charged pions from the FOPI experiments~\cite{FOPIb} as well as the corresponding theoretical simulations in the same reactions with and without the consideration of SRCs. It is obvious to see that the theoretical simulations of $^{197}$Au + $^{197}$Au collisions using the HMT-SCGF parameter can indeed reproduce the \rpi ratio of the FOPI data fairly under the symmetry energy setting in a range with parameter $x$ no less than 1. Therefore, from the consistences of both pions multiplicities and their ratios in theoretical simulations using the HMT-SCGF parameter with the FOPI experimental data, we can conclude that this isospin-dependent single NMD given in Eq.~(\ref{nk}) as well as the corresponding HMT-SCGF parameter are reliable for the IANM. Correspondingly, this result also qualitatively favors a mildly soft prediction for the nuclear symmetry energy at high densities. Certainly, one can also observe that without the consideration of SRCs the ratio \rpi of charged pions can also reproduce the FOPI experimental data even with all $x$ parameters used here. However, it needs to be emphasized that the multiplicities of both $\pi^{-}$ and $\pi^{+}$ in this case deviate too much from the FOPI experimental data.

\begin{figure}[t]
	\centerline{\includegraphics[width=0.9\columnwidth]{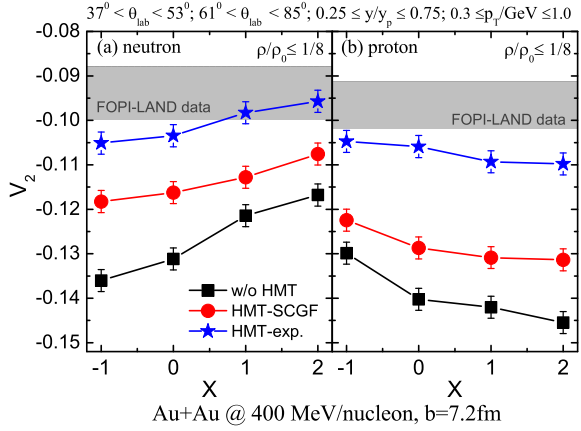}}
	\caption{(Color online) Elliptic flows of neutrons (a) and protons (b) compared with those from the FOPI-LAND experiment in semicentral $^{197}$Au + $^{197}$Au collisions with a impact parameter $b=7.2~$fm at 400 MeV/nucleon with and without the consideration of SRCs. } \label{v2}
\end{figure}
\begin{figure}[t]
	\centerline{\includegraphics[width=0.9\columnwidth]{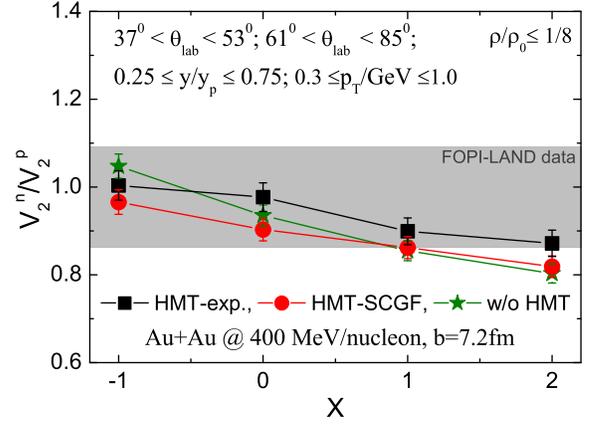}}
	\caption{(Color online) Ratios of neutron elliptic flows over proton elliptic flows compared with those from the FOPI-LAND experiment in semicentral $^{197}$Au + $^{197}$Au collisions with a impact parameter $b = 7.2$ fm at 400 MeV/nucleon with and without the consideration of SRCs. } \label{v2r}
\end{figure}

We now investigate effects of the SRCs on the collective flows of nucleons. As a typical collective flow, the elliptic flows of nucleons defined as
\begin{equation}
v_{2}={\big\langle}\frac{p^{2}_{x}-p^{2}_{y}}{p^{2}_{x}+p^{2}_{y}}{\big\rangle},
\end{equation}
are widely used as the probe of the isovector part of nuclear interactions in HICs at intermediate energies as well as the properties of the hot and dense matter formed in the early stage of HICs at relativistic energies, see, e.g., Refs.~\cite{Dan02,LiPC99,Song11}. Presently, predictions on the nuclear symmetry energy using the elliptic flow of nucleons are mainly through comparing theoretical simulations with the available data in FOPI-LAND and/or ASY-EOS experiments~\cite{FOPI-LAND,ASY-EOS}. For example, based on these data, the T\"{u}bingen quantum-molecular-dynamics (T\"{u}QMD)~\cite{Cozma13} model and ultrarelativistic quantum-molecular-dynamics (UrQMD)~\cite{Trau12} model consistently favor a moderately soft prediction for the nuclear symmetry energy at high densities. Nevertheless, to extract the precise information about the nuclear symmetry energy using the elliptic flows of nucleons in HICs, it needs to know how the SRCs affect the elliptic flows of nucleons. On the other hand, to further check the reliability of this isospin-dependent NMD as well as the corresponding parameter settings, we compare our theoretical simulations of elliptic flows of nucleons with those from the experiment in FOPI-LAND collaboration~\cite{FOPI-LAND,Cozma13} during examining effects of the SRCs on the elliptic flows of nucleons. Shown in Fig.~\ref{v2} are the elliptic flows of nucleons in semicentral $^{197}$Au + $^{197}$Au collisions with a impact parmater $b=7.2~$fm and a beam energy of 400 MeV/nucleon. Obviously, with the consideration of SRCs, the squeezed out elliptic flows of both neutrons and protons are decreased due to the correlations enhance the difficulties of anisotropic nucleons emissions, and this phenomenon is especially apparent for the scenario using the HMT-exp. parameter because of more strong correlation effects in this scenario. Again, because the dominance of $np$ SRC pairs gets the isospin asymmetry of reaction systems to be reduced, we can also observe a reduced sensitivity of elliptic flows to the nuclear symmetry energy. Moreover, it can be seen that the elliptic flows of protons in FOPI-LAND experiment are failed to reproduce in our theoretical simulations regardless of using the HMT-SCGF parameter or the HMT-exp. parameter. Interestingly, however, it is found that with the symmetry energy setting in a range with parameter $x$ no less than about 0.9, the elliptic flows of neutrons in FOPI-LAND experiment can be well reproduced by theoretical simulations of $^{197}$Au + $^{197}$Au collisions employing the HMT-exp. parameter, this constraint on the symmetry energy coincides qualitatively with the previous predictions using the T\"{u}QMD~\cite{Cozma13} and UrQMD~\cite{Trau12} models. Consequently, it can be concluded from this result that the HMT-exp. parameter should not be ruled out roughly at present. Finally, similar to the ratio \rpi of charged pions, the ratio of the neutron elliptic flow over the proton elliptic flow $v_{2}^{n}/v_{2}^{p}$ is also an effective probe in probing the symmetry energy in HICs. Therefore, we show in Fig.~\ref{v2r} this ratio in the same reaction. It can be seen that, setting the symmetry energy in a range with parameter $x$ approximately no more than 1, the ratio $v_{2}^{n}/v_{2}^{p}$ in FOPI-LAND experiment can be well reproduced by theoretical simulations of $^{197}$Au + $^{197}$Au collisions in all scenarios with or without the consideration of SRCs. Moreover, even for all $x$ parameters used here, the theoretical simulations of $^{197}$Au + $^{197}$Au collisions using the HMT-exp. parameter can well reproduce this ratio of the FOPI-LAND experimental. This implies again that we should not roughly rule out the HMT-exp. parameter at present. Nevertheless, considering that the elliptic flows of protons in these cases do not fit the FOPI-LAND experimental data,
we naturally can not put stringent constraints on the high-density symmetry energy according to we have learned from the elliptic flow signals in this study.

So far, according to what we have learned from the elliptic flow signals and taking the above obtained from the pion signals an overall, we can conclude two points. First, the parameterized isospin-dependent single NMD given in Eq.~(\ref{nk}) is reliable for the IANM, and both the HMT-SCGF parameter and the HMT-exp. parameter can not be ruled out according to the available experimental information at present. Second, according to the pion signals of theoretical simulations using the HMT-SCGF parameter as well as their comparisons with the FOPI experimental data, we indeed can make a mildly soft conclusion for the nuclear symmetry energy at high densities, it is however, our theoretical simulations of $^{197}$Au + $^{197}$Au collisions using both HMT-SCGF and HMT-exp. parameters can not fit the experimental pion data and the experimental elliptic flow data simultaneously, we therefore can not put stringent constraints on the high-density symmetry energy. 

\section{Summary}\label{Summary}
Within an IBUU transport model using as the input nucleon momentum profiles from an isospin-dependent single NMD with a HMT induced by SRCs, in this paper we have investigated effects of the SRCs on the pion and flow observables in $^{197}$Au + $^{197}$Au collisions at 400 MeV/nucleon. Compared to the case without the SRCs, the SRCs are found to cause more increment of $\pi^{+}$ than $\pi^{-}$ in $^{197}$Au + $^{197}$Au collisions, leading to an appreciable reduction of the \rpi ratio; while for the flow observables, the SRCs are found to decrease the squeezed out elliptic flows of both neutrons and protons due to the correlations enhance the difficulties of anisotropic emission of nucleons. Moreover, because of the dominance of $np$ SRC pairs in the HMT and thus a corresponding reduction of the isospin asymmetry of reaction systems, both pion and flow observables are found to show a reduced sensitivity to the nuclear symmetry energy. On the other hand, through comparing the pion observable as well as the flow observable with the available data, we have also checked the reliability of the used isospin-dependent single NMD as well as the corresponding two sets of parameters. It is found that the theoretical simulation of $^{197}$Au + $^{197}$Au collisions using the HMT-exp. parameter and the HMT-SCGF parameter, respectively, can reproduce the neutron elliptic flows of the FOPI-LAND experiment and the \rpi ratios of the FOPI experiment under the symmetry energy setting in a certain range.
Therefore, we can conclude that this parameterized isospin-dependent single NMD is reliable for the IANM, and both the HMT-SCGF parameter and the HMT-exp. parameter can not be ruled out roughly according to the available experimental information at present. On the contrary, because our theoretical simulations of $^{197}$Au + $^{197}$Au collisions can not fit the experimental pion data and the experimental elliptic flow data simultaneously, we therefore can not put stringent constraints on the high-density symmetry energy. Nevertheless, our studies explicitly indicate that the cross examinations of various observables using various experimental data are the necessary solution to the determination of nuclear symmetry energy, and the symmetry energy measurement experiment at RIBF-RIKEN in Japan~\cite{SEP} using the SAMURAI-Time-Project-Chamber~\cite{Shane15} may throw more light on the determination of nuclear symmetry energy at high densities.

\begin{acknowledgments}
G. F. Wei would like to thank Profs. Gao-Chan Yong and Bao-An Li for helpful discussion. This work is supported by the National Natural Science Foundation of China under grant Nos.11965008, 11405128, U1731218, and Guizhou Provincial Science and Technology Foundation under Grant No.[2020]1Y034, and the PhD-funded project of Guizhou Normal university (Grant No.GZNUD[2018]11),  and also in part by the Xi'an Science and Technology Planning project under grant No.CXY1531WL35.
\end{acknowledgments}

\end{document}